\documentclass[floatfix,showpacs,pra,twocolumn,aps]{revtex4}

 \usepackage{bm}
 \usepackage{amsmath}
 \usepackage{amssymb}
 \usepackage{latexsym} 
 \usepackage{amsfonts} 
 \usepackage{epsfig}
 \usepackage{psfrag}

 \newcommand{\f}[1]{\mbox{\boldmath$#1$}}

 \newcommand{\ket}[1]{|#1\rangle} 
 \newcommand{\bra}[1]{\langle#1|} 
 \newcommand{\bracket}[2]
 {\langle#1|#2\rangle} 
 \newcommand{\nn}{\nonumber\\} 
 \newcommand{\bea}{\begin{eqnarray}}
 \newcommand{\ea}{\end{eqnarray}}
 \newcommand{\eea}{\end{eqnarray}}
 \newcommand{\order}[1]{{\cal O}\{{#1}\}}
 \newcommand{\ord}{{\cal O}}

\newcommand{\abs}[1]{\left|#1\right|}

\begin{document}

\title{Quantum algorithm for optical template recognition with noise
  filtering}   

\author{Gernot Schaller and Ralf Sch\"utzhold$^*$}

\affiliation{Institut f\"ur Theoretische Physik, 
Technische Universit\"at Dresden, 
D-01062 Dresden, Germany}

$^*$ email: {\tt schuetz@theory.phy.tu-dresden.de}

\begin{abstract} 
We propose a probabilistic quantum algorithm that decides whether a 
monochrome picture matches a given template 
(or one out of a set of templates).
As a major advantage to classical pattern recognition, the algorithm
just requires a few incident photons and is thus suitable for very
sensitive pictures (similar to the Elitzur-Vaidman problem).
Furthermore, for a $2^{n}\times 2^{m}$ image, $\ord(n+m)$ qubits
are sufficient. 
Using the quantum Fourier transform, it is possible to improve the 
fault tolerance of the quantum algorithm by filtering out small-scale
noise in the picture.
For example images with $512\times512$ pixels, we have numerically
simulated the unitary operations in order to demonstrate the
applicability of the algorithm and to analyze its fault tolerance. 
\end{abstract} 


\maketitle

\section{Introduction}\label{Sintroduction}

It is well known that quantum computers are suited to solving certain 
classes of problems much better than classical computers.
A prominent example is Shor's algorithm \cite{shor1997} for factoring 
an integer number with an effort that grows polynomially in the number
of its digits, which is believed to be classically impossible.
A further impressive example is Grovers algorithm \cite{grover1997} 
for finding an item in an unsorted database: 
There, the effort grows only as the square-root $\sqrt{N}$ of the
number $N$ of database entries on a quantum computer, whereas it grows 
with $N$ on a classical computer. 
In addition, there are further black box problems such as the Deutsch
\cite{deutsch1985} and the Deutsch-Josza algorithm
\cite{deutsch1992} as well as others
\cite{elitzur1993,simon1997,bernstein1997} 
(for an overview see e.g., \cite{nielsen2000}). 

The possible speedup of quantum algorithms is essentially enabled by 
the feature of quantum parallelism. 
This parallelism permits to calculate with a superposition of states
on a quantum computer, which is not possible on classical computers. 
The first quantum computers have already been constructed. 
For example, Shor's algorithm has been used on an NMR quantum computer
\cite{vandersypen2001} to factorize the number 15. 
This is certainly not impressive if one considers the smallness of the
number but nevertheless serves as a proof of principle. 

However, there exists a plethora of further classically challenging 
problems such as e.g., pattern recognition \cite{fukunaga1972}, which
can also benefit from the application of quantum algorithms
\cite{horn2002,sasaki,trugenberger}.
It has already been demonstrated that quantum parallelism can be 
exploited to identify and localize a regular simple pattern
\cite{schuetzhold2003} within an otherwise unstructured picture. 
Here, we will present a probabilistic quantum algorithm that is
capable of recognizing an arbitrary image 
(even in the presence of noise at a moderate level) 
after shining a few photons on the picture.
In contrast to previous pattern matching approaches 
(see, e.g., \cite{trugenberger}), it does not require the copying of
quantum states 
(which is only possible probabilistically for general states) and
enables a (probabilistically) non-destructive measurement,
cf.~\cite{elitzur1993}.  

\section{Problem Definition}\label{Sproblem_definition}

Following the problem description of \cite{schuetzhold2003}, let us 
consider a large rectangular $N_x \times N_y$ array of unit cells that
may either be black (absorptive) or white (reflective). 
To allow for a binary representation, we will consider cases where the 
number of array cells in every dimension are powers of 2, i.e.,
$\log_2 N_{x/y}=n_{x/y}$ with integers $n_{x/y}$. 
Further-on, we will denote the pixels in the array that are reflective
(white) as {\bf points}.

The problem consists of recognizing whether the pattern in the array 
matches a given template (for example, a letter of an alphabet). 
Note that this slightly differs from template matching discussed
e.g., in \cite{trugenberger,sasaki}, where a template is to be 
found that optimally matches the given unknown quantum state. 
With using varying templates however, one can evidently establish a 
relation between these problems.
Some aspects of the presented quantum algorithm will be similar to the 
discrimination problem of known quantum states \cite{bergou2004} via
suitably chosen positive operator valued measures (POVM), which will
be discussed in section \ref{Sdiscrimination}.

The classical approach to measure the displayed pattern on the array 
would be to shine light 
(consisting of many, say $\order{N_x \times N_y}$, photons) on the
array and to measure absorption and transmission accordingly.
However, in the case we consider here, the array is also assumed to be
very sensitive (imagine, for example, an exposed but not yet developed
film or a pattern of partially fluorescent ions in a Paul trap), 
such that each absorbed photon causes a certain amount of damage. 
Evidently, the classical measurement approach would significantly
disturb the system. 

One might worry that the momentum of the reflected photon (recoil)
also disturbs the picture, but this effect will be extremely small if
the mass of a pixel is sufficiently big.
Alternatively, one might imagine the reflective pixels to be
transparent and to place a mirror behind the array.

Then, a quantum algorithm can cope with such a task: 
The feature of quantum parallelism can be exploited by storing the
relevant information of the image in a quantum superposition state
with using just a single photon and not (or only very little)
destroying the image.
The task is thus similar to the Elitzur-Vaidman problem
\cite{elitzur1993}, which allows for testing for the existence of an
object without any energy-momentum exchange.

\section{Read-Out Scheme}\label{Sread-Out}

Generally, the coordinates $(x,y)$ of a pixel in the image can be 
written in their binary representation
\bea
x={x_1}\otimes\ldots\otimes{x_{n_x}} 
&\qquad&
y={y_1}\otimes\ldots\otimes{y_{n_y}}\,,
\eea
where $x_i$ and $y_j$ denote the $i$-th and $j$-th bit of $x$ and $y$, 
respectively, i.e., \mbox{$x=x_12^{n_x-1}+x_22^{n_x-2}+\ldots+x_{n_x}$}. 
Regarding these coordinates $(x,y)$ as control qubits 
\mbox{$\ket{x}=\ket{x_1}\ket{x_2}\ldots\ket{x_{n_x}}$}, the photon probing
the image can be entangled with the coordinates in the following way: 

The photon passes through a series of quantum controlled refractors  
$R_{n_x}^x \ldots R_1^x R_{n_y}^y \ldots R_{1}^y$ 
that effectively displace the photon by defined distances 
$\Delta x_i$ or $\Delta y_i$, if the control qubit $\ket{x_i}$ or
$\ket{y_i}$ is in the state $\ket{1}$ and leave it unaffected if the
control bit is in the state $\ket{0}$.
By choosing the displacement of the refractor $R_i^{x}$ as 
$\Delta x_i = 2^i \Delta x_0$ (and likewise for the other refractors
$R_i^y$), the final displacement of the 
photon corresponds to the coordinates $(x,y)$ which encode the
position of a single image pixel.

\begin{figure}[ht]
\includegraphics[width=5.3cm]{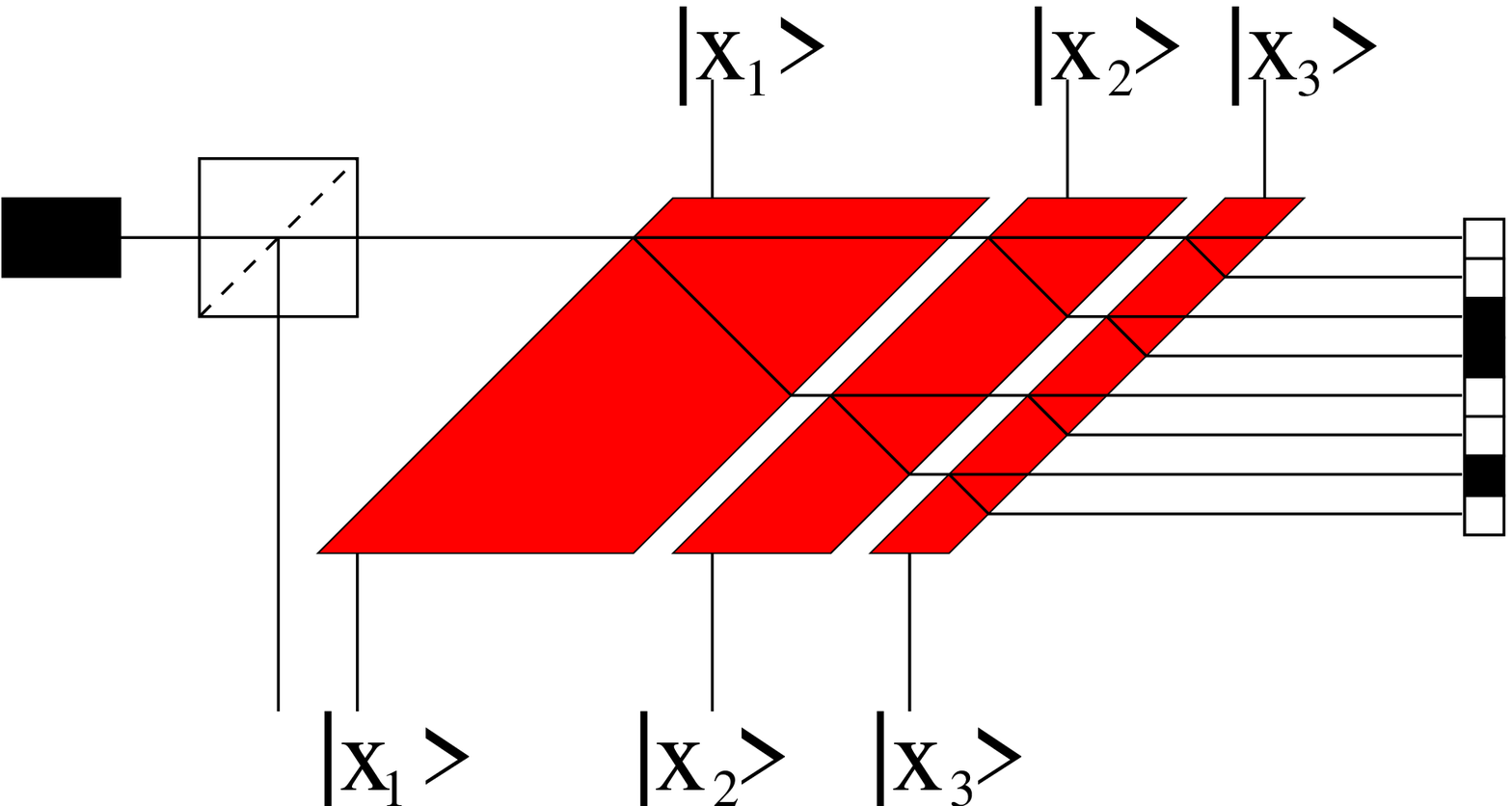}
\includegraphics[width=3.2cm]{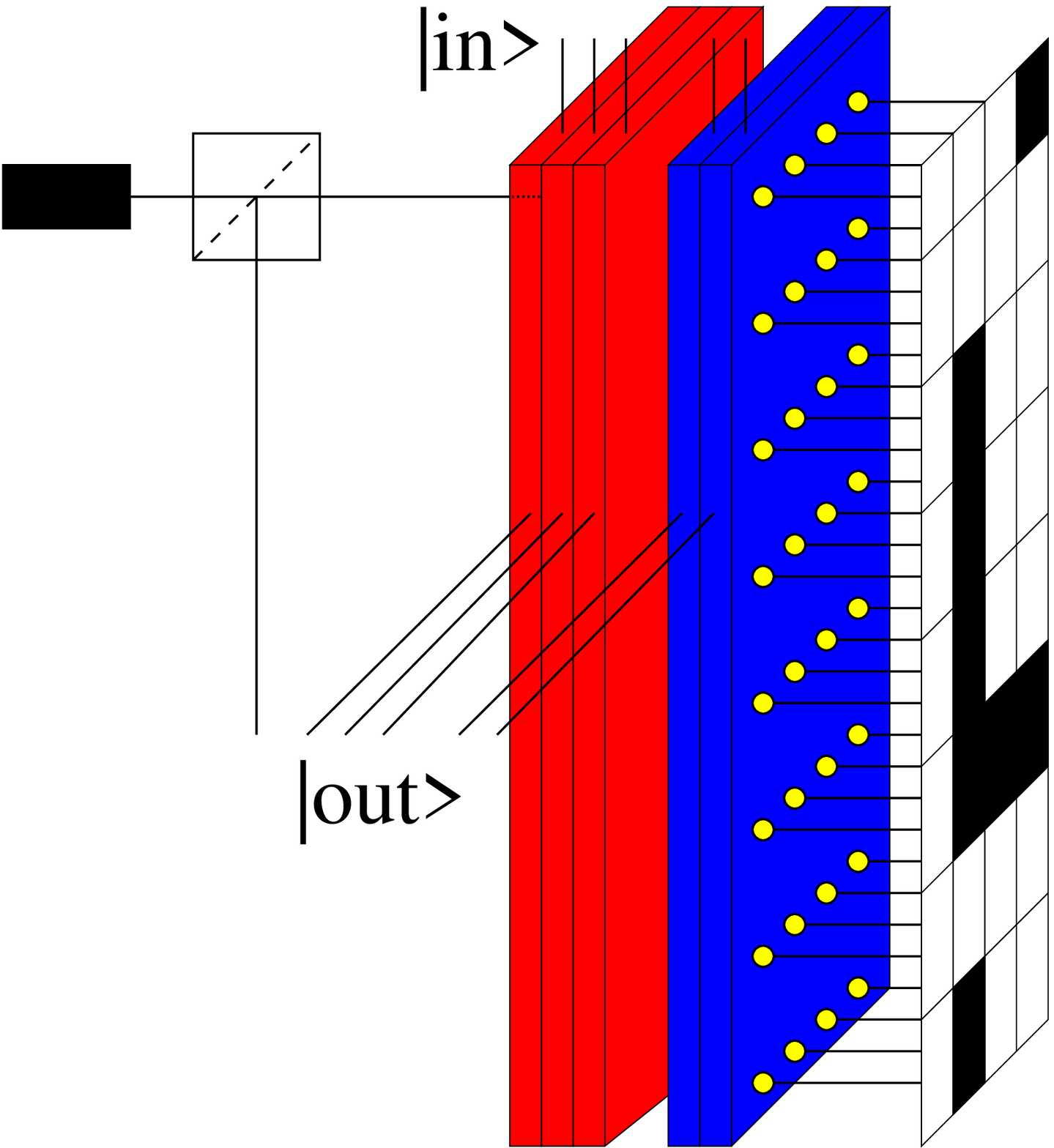}
\caption{\label{Fpreparation}
[Color Online]
{\bf Left}: 
Illustration of a series of three quantum-controlled refractors
required for an 8-pixel one-dimensional array. 
A photon created by the source (box in the left) passes the refractor
series, interacts with the pixel array and takes -- if reflected --
the reverse path. 
Depending on the value of the control qubits $\ket{x_i}$, the
refraction index of the medium is changed, which effectively produces
a displacement of the photon. 
{\bf Right}: 
Schematic representation of the two-dimensional configuration 
(exemplified for an $8 \times 4$ array). 
The first three refractors generate the displacement in
$x$-direction (for 8 pixels) and the remaining two act in the 
$y$-direction (for 4 pixels).}
\end{figure}

In a laboratory setup, this could for example be realized by using a
varying refractor thickness, see figure \ref{Fpreparation} left panel. 
For quantum controlled devices, one can generate superposition
states of several pixel positions in the image as control qubits.
Storing these bits in a coherent superposition state containing all
pixels with equal amplitudes, a single photon can be forced to
interact simultaneously and uniformly with the complete array.

One should be aware that this causes a strong entanglement between the 
photon position and the control qubits.
Thus, any measurement on the photon would affect the refractor control 
qubits as well and completely decohere them. 
The interaction of the photon with the image corresponds to absorption  
or reflection. 
In case of (perpendicular) reflection, the photon will reverse its
original path (this is enforced by the entanglement with the refractor
control qubits) and after passing through all refractors again, all
information about the way of the photon (given that it was reflected)
is lost.
Thus, the entanglement with the control qubits is partially removed 
(quantum eraser) and we obtain a coherent superposition of the
reflecting pixels (points). 
In the other case (absorption), the entanglement cannot be removed,
and the algorithm fails -- i.e., one has to send another photon if
permitted by the fragility of the quantum array. 
As a formal simplification, one can consider the action of the
configuration of the setup in figure \ref{Fpreparation} as a quantum
black box: 
\bea
{\mathcal B} 
\;:\; 
\ket{x}\otimes
\ket{y}\otimes
\ket{0}
\,\to\,
\ket{x}\otimes
\ket{y}\otimes
\ket{f(x,y)}\,,
\eea
which encodes the output in the characteristic function $f(x,y)$ of
the image. 
The function $f(x,y)$ takes the value 1 if the pixel $x \otimes y$ is
reflective (i.e., if $\ket{x}\otimes\ket{y}$ encodes a point) and 0
if the pixel is black (i.e., if the photon is absorbed).
If the control qubits on the refractors are initially prepared in a
superposition state (e.g., acting Hadamard gates on each qubit),  
the characteristic function $f(x,y)$ of the image is tested for all
pixels simultaneously 
\bea
{\mathcal B}\left[
{\mathcal H}^{(n_x)}\ket{0^{(n_x)}}\otimes
{\mathcal H}^{(n_y)}\ket{0^{(n_y)}}\otimes
\ket{0}
\right]
=\nn
\frac{1}{\sqrt{N_x N_y}}
\sum_{x=0}^{N_x-1}
\sum_{y=0}^{N_y-1}
\ket{x}\otimes
\ket{y}\otimes
\ket{f(x,y)}\,.
\eea
Measuring the third register 
(i.e., the existence of a reflected photon) 
and obtaining $\ket{1}$ as a result prepares the quantum state
as an uniform superposition of all points in the image. 
The other outcome ($\ket{0}$) corresponds to the absorption of the
photon and would lead to entanglement between the refractor control
qubits and the image.
Thus, with the outcome $\ket{1}$ (outgoing photon) one has prepared a
quantum state (in the refractor control qubits) that is suitable for
performing further calculations. 
An alternative scheme based on linear optics, which could be applied
if the image is not too large, is discussed in the Appendix.
Note that this scheme has the advantage that the image does not have 
to be loaded into a possibly fragile quantum memory
(cf.~\cite{trugenberger,sasaki}). 

\section{Quantum Algorithm}\label{Squantum_algorithm}

After extracting the superposition state containing all points of the
image
\bea
\label{points}
\ket{\Psi}
=
\sum_{x=0}^{N_x-1}
\sum_{y=0}^{N_y-1}
\frac{f(x,y)}{\sqrt{M}}
\ket{x}\ket{y}
=
\sum\limits_{(x,y)\in\rm image}
\frac{\ket{x,y}}{\sqrt{M}}
\,,
\ea
where $M$ with $1 \ll M < N_x N_y$ denotes the total number of points, 
one has to decide whether it corresponds to a given template. 

Obviously, a $N_x \times N_y$ array (in black and white) could contain
$2^{N_x N_y}$ different images -- but the Hilbert space of all possible
quantum state has merely $N_x N_y=2^{n_x+n_y}$ dimensions.
Hence different images will not correspond to orthogonal quantum
states in general and thus it is -- even in the absence of noise --
not possible to distinguish them with certainty, i.e., the presented
quantum algorithm can only work probabilistically.

The problem of efficiently discriminating non-orthogonal quantum
states has been studied extensively and is usually formulated within
the framework of positive operator valued measures (POVM):
Given a quantum state out of a set of perfectly known (non-orthogonal) 
quantum states, it is possible to construct a POVM which decides
probabilistically which one of these states it is with minimum 
error probability, see section \ref{Sdiscrimination}. 
However, the design of a suitable POVM will be -- especially in the  
case of large images (high-dimensional Hilbert-spaces) with more than
two possible quantum states -- quite demanding in general  
\cite{bergou2004,ariano2005}. 
In addition, if the quantum states are not known exactly -- e.g., in
the presence of perturbations and noise -- the construction of the
optimal POVM is even more complicated.
Finally, the physical realization of such an optimal POVM will also be
quite demanding in general and can easily destroy part of the speed-up
of the quantum algorithm.  
Therefore, we propose a different procedure that shifts the effort
from designing a suitable POVM towards the implementation of a quantum 
oracle gate (in an inverse application of Grovers algorithm). 

Using the characteristic function $f(x,y)$ of the image (which yields
1 for the marked items, i.e., points of the image, and 0 otherwise) 
as an oracle, a Grover iteration generates a rotation in the 
two-dimensional sub-space of the Hilbert space spanned by the coherent
superpositions of all numbers 
\mbox{$\ket{s}={\mathcal H}^{(n_x)}\ket{0^{(n_x)}}\otimes
{\mathcal H}^{(n_y)}\ket{0^{(n_y)}}$} 
on the one hand and the numbers of the marked items, i.e.,  
the state $\ket{\Psi}$ in the above equation, on the other hand. 
Since there are $M$ solutions to this search problem,
\bea
R=\left\lceil{\frac{\pi}{4}\sqrt{\frac{N_x N_y}{M}}}\right\rceil
\ea
inverse Grover iterations will rotate the original state $\ket{\Psi}$
in Eq.~(\ref{points}) into the total superposition state $\ket{s}$, 
or at least close to it 
(discretization error, cf.~\cite{discretization}).
After application of a Hadamard gate to each of the qubits, the
resulting state would be $\ket{0^{(n_x)}}\otimes\ket{0^{(n_y)}}$.
If the image indeed perfectly equals the template -- i.e., if their
characteristic functions coincide -- and if the Grover iterations were
assumed to be perfect (no discretization error),  the state
$\ket{\Psi}$ in Eq.~(\ref{points}) prepared after measuring the
returning photon would be unitarily transformed into the final state 
$\ket{0^{(n_x)}}\otimes\ket{0^{(n_y)}}$.
Consequently, a measurement in the computational basis would yield
zeros for all bits.

If the image differs from the template 
(and assuming a perfect Grover rotation \cite{discretization}), 
the probability of ending up in the final state 
$\ket{0^{(n_x)}}\otimes\ket{0^{(n_y)}}$ would be given by the overlap
of the characteristic functions of the template $f(x,y)$ and the image
$f'(x,y)$ 
\bea
p=\frac{1}{MM'}
\left|
\sum_{x=0}^{N_x-1}
\sum_{y=0}^{N_y-1}
f(x,y)f'(x,y)
\right|^2
\ea
since all involved operations were unitary and hence probability
conserving.
Since this overlap determines how similar image and template are,
measuring all qubits at the end and obtaining zero everywhere is a
strong indication that the image equals the template or at least is
very similar to it.
Obtaining 1 somewhere, on the other hand, indicates that the image
does not equal the template with high probability
\cite{discretization}. 
As usual, the result can be made more decisive by repeating the whole
algorithm.

\section{Noise Filtering}\label{Snoise_filtering}

So far, the input state $\ket{\Psi}$ was assumed to be perfect.
However, in reality neither the realization of the image in the array
nor the reflection and absorption properties of the image pixels can
be assumed as perfect. 
Therefore, the basic algorithm described above can be improved using
the quantum Fourier transform to reduce possible perturbations: 
As in classical pattern recognition, one can perform a cutoff of large  
wave-numbers, which reduces noise such as pixel defects.
In a quantum algorithm, such a cutoff has to be implemented in a
suitable way -- just measuring the $k$-value and subsequently deciding
whether it lies above of below the cutoff does not work, since a hard
$k$-measurement completely decoheres the quantum state.
Therefore, a noise-filter should be realized via the measurement of an
ancilla qubit which has been coupled to the control qubits:
To this end, we introduce a unitary noise-filter operator 
\bea
{\mathcal N}
\left[\ket{k_x}\otimes\ket{k_y}\otimes\ket{0}\right] =
\nn
\ket{k_x}\otimes\ket{k_y}\otimes\left\{\cos[\vartheta(k_x,k_y)]\ket{1}
+\sin[\vartheta(k_x, k_y)]\ket{0}\right\}
\,,
\ea
with a suitably chosen noise-filter function $\vartheta(k_x, k_y)$. 
Measuring the last register and obtaining $\ket{1}$ performs the
desired cutoff in frequency space.

\begin{figure}[ht]
\includegraphics[width=8.8cm]{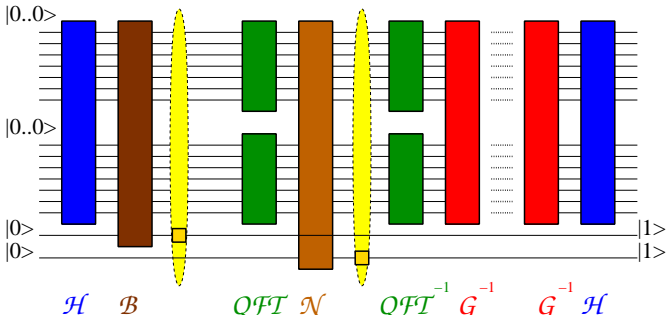}
\caption{\label{Fcircuit}
[Color Online]
Quantum circuit of the template matching algorithm for 14 control
qubits (7 for $x$ and $y$ respectively) plus 2 ancilla qubits
(ancilla qubit values shown for a successful run).}
\end{figure}

Let $M_{\rm tp}$ denote the number of template points and 
$M_{\rm im}\approx M_{\rm tp}$ the number of image points.
Then, the quantum algorithm can be summarized as follows 
(compare also figure \ref{Fcircuit}):
\begin{enumerate}
\item 
initialize the state vector with uniform superpositions in $x$ and $y$
as well as two ancilla qubits\\ 
${\ket{\Psi_1}=\mathcal H}^{(n_x)}\ket{0^{n_x}} \otimes {\mathcal
H}^{(n_y)}\ket{0^{n_y}} \otimes \ket{0} \otimes \ket{0}$
\item 
apply the black box with the first ancilla qubit\\
${\ket{\Psi_2}={\mathcal B}\ket{\Psi_1}}$
\item 
measure the first ancilla qubit and proceed after obtaining the
outcome $\ket{1}$\\
$\ket{\Psi_3}\stackrel{\ket{1}}{=}M_{\rm im}^{-1/2} 
\sum\limits_{(x,y)\in\rm image}
\ket{x}\otimes\ket{y}\otimes\ket{1}\otimes \ket{0}$
\item 
perform quantum Fourier transform in $\ket{x}$ and $\ket{y}$\\
$\ket{\Psi_4}={\cal QFT}_{x,y}\ket{\Psi_3}$
\item 
apply noise filter operator with second ancilla\\
$\ket{\Psi_5}={\mathcal N}\ket{\Psi_4}$
\item
measure the second ancilla qubit and proceed after obtaining the
outcome $\ket{1}$\\
$\ket{\Psi_6}\stackrel{\ket{1}}{=}
\sum_{k_x,k_y}
\tilde f_{\rm cut}(k_x,k_y)
\ket{k_x}\otimes\ket{k_y}\otimes\ket{1}\otimes\ket{1}$
\\
with 
$\tilde f_{\rm cut}(k_x,k_y)=\tilde f(k_x,k_y)\cos[\vartheta(k_x,k_y)]$
\item
perform the inverse quantum Fourier transform\\
$\ket{\Psi_7}={\cal QFT}_{x,y}^\dagger\ket{\Psi_6}$
\item 
perform 
$R=\left\lceil{\frac{\pi}{4}\sqrt{\frac{N_x N_y}{M_{\rm tp}}}}\right\rceil$ 
inverse Grover iterations\\ 
$\ket{\Psi_8}={\cal G}^{-R}\ket{\Psi_7}$
\item 
apply Hadamard gates on non-ancilla qubits\\
$\ket{\Psi_9}={\mathcal H}^{(n_x)}\otimes{\mathcal H}^{(n_y)}\ket{\Psi_8}$
\item 
measure final state in computational basis\\
$\ket{\Psi_9}\stackrel{?}{=}
\ket{0^{(n_x)}}\otimes\ket{0^{(n_y)}}\otimes\ket{1}\otimes\ket{1}$
\end{enumerate}
The algorithm fails when measurement of one of the ancilla qubits
yields $\ket{0}$, i.e., if the photon is absorbed or if a projection
onto the wrong $k$-values is performed.

The probability of obtaining finding the system in the state
$\ket{0\ldots 0}\otimes\ket{1}\otimes\ket{1}$ as the result of the last
measurement corresponds to the probability of template matching, i.e.,
given a template such as the letter {\bf A} and an initial state such
as the (noise-perturbed) state $\ket{B}$, the quantum algorithm
decides whether the template is matched or not \cite{discretization}. 

\section{Algorithmic Performance}\label{Salgorithmic_performance}

\begin{figure}[ht]
\includegraphics[width=4.2cm]{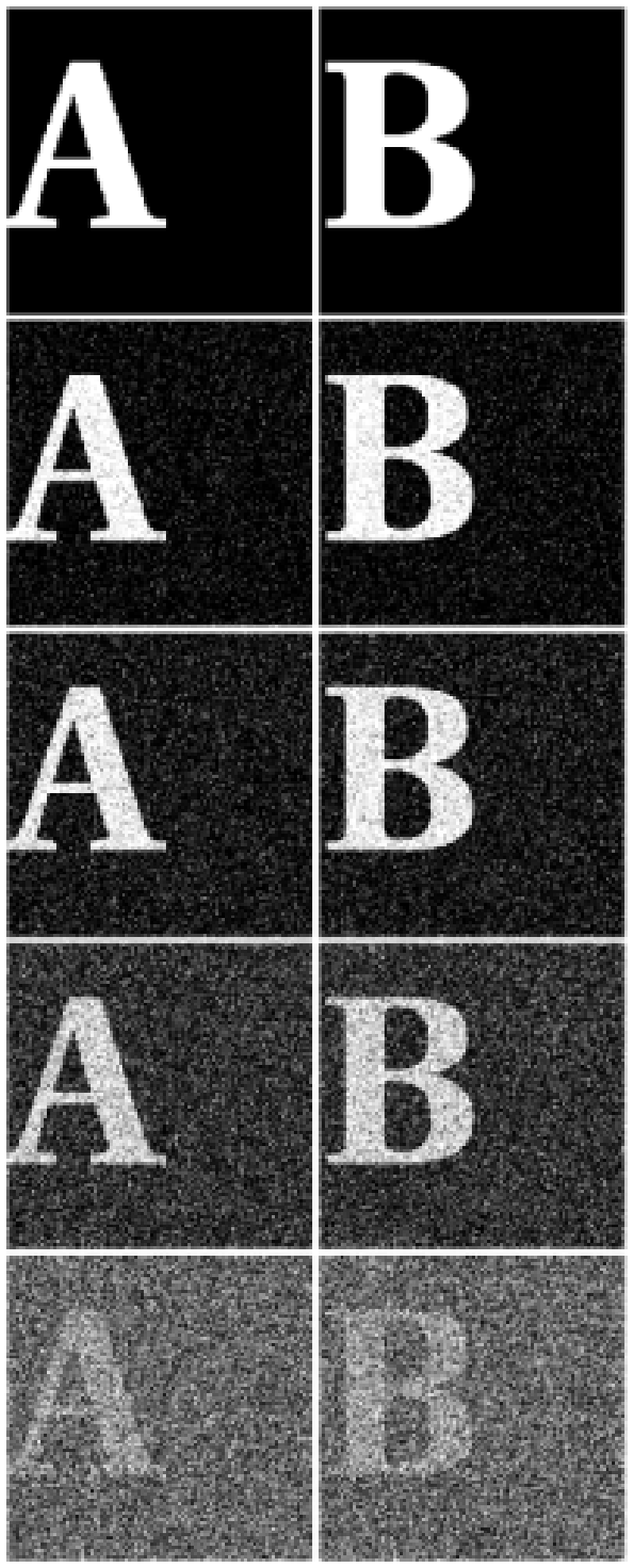}
\includegraphics[width=4.2cm]{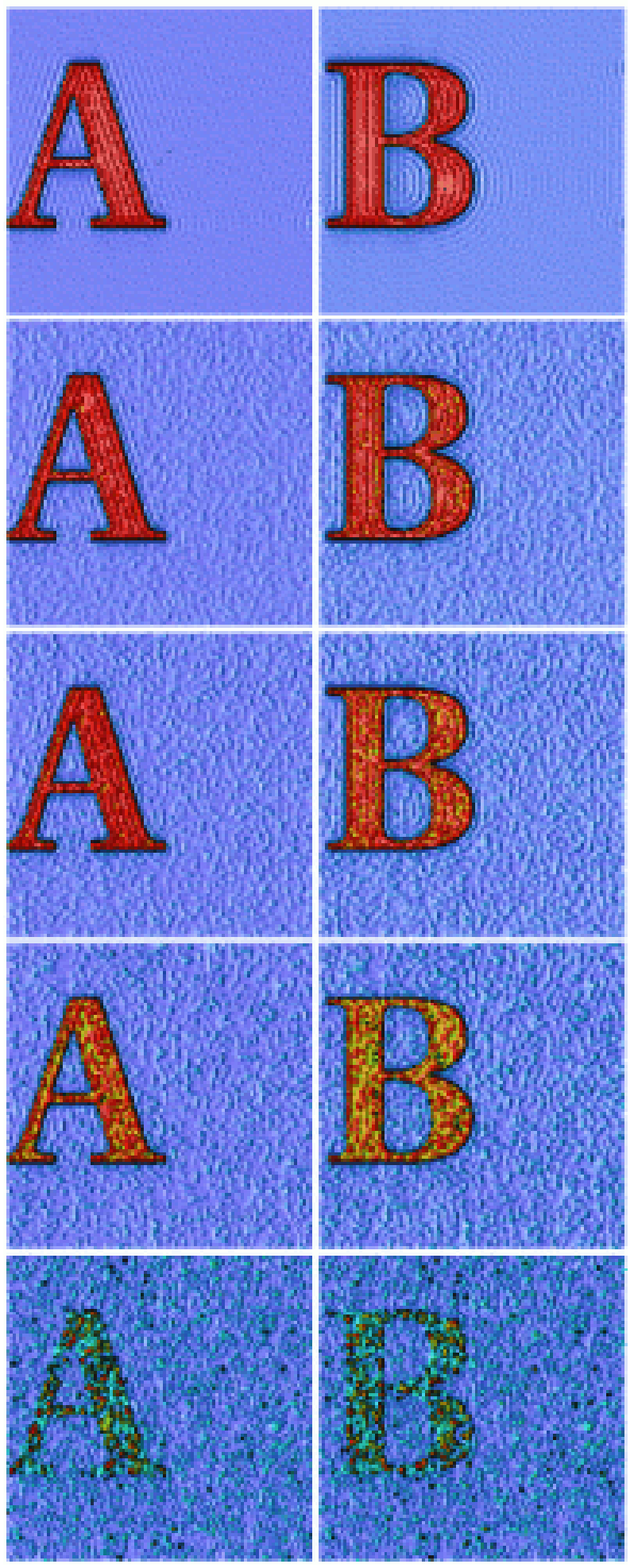}
\caption{\label{Fimage}
[Color Online]
{\bf Left}: 
Images of the input states ($512\times 512$) with various noise
levels. 
From top to bottom the image pixels of the initial state
have been inverted with a probability of 
0\%, 5\%, 10\%, 20\%, and 40\%, respectively.
{\bf Right}: 
Panels show the distribution of the squared amplitude of the inverse of the
Fourier-smoothed images 
(i.e., after step 6 of the quantum algorithm). 
Only $k$-values with $0 < k < 40$ have been kept. 
All amplitudes larger than half the maximum amplitude in every image
are shown in red, whereas the thin black isoline encodes the quarter
of the maximum amplitude.
}
\end{figure}

Starting from a possibly noise-perturbed initial state as prepared by
step 3 of the quantum algorithm, we have numerically simulated the
action of the corresponding unitary gates and measurements for the 
$512\times512$ example arrays in figure \ref{Fimage}.
As the number of allowed states increases exponentially with the
number of simulated $n_x+n_y$ qubits, the numerical simulations would 
involve $2^{n_x+n_y} \times 2^{n_x+n_y}$ matrices which do not fit into the
main memory. 
Fortunately, the involved unitary operations can be expanded into
combinations of one or two qubit operations \cite{nielsen2000}, 
which can be calculated.
Thus, the whole algorithm for pattern recognition for $n_x+n_y=18$
qubits runs in a time of few seconds.
As a noise-filter function a sharp cutoff was used, i.e., 
\bea
\vartheta(k_x, k_y) = \left\{
\begin{array}{ccc}
0 & : & 0 < \sqrt{k_x^2 + k_y^2} < k_{\rm max}
\\
\pi/2 & : & \mbox{otherwise}
\end{array}
\right.
\,,
\eea
which leads to a simple projection on the allowed $k$-values. 
In addition to the high-frequency ($k \ge k_{\rm max}$) components, 
it can be advantageous to remove the \mbox{$k_x=k_y=0$} component as well,
especially if the noise significantly changes the total number of
points in the image. 

The computational complexity of the algorithm depends on the total
number of pixels and on the number of points in the template.
The Hadamard gates and the quantum controlled refractors require 
$\order{n_x+n_y}$ operations and the quantum Fourier transforms
involve $\order{n_x^2+n_y^2}$ gates.
The necessary number of Grover iterations depends on the size of the
template \mbox{$R=\order{\sqrt{N_xN_y/M_{\rm tp}}}$}, and the number of
involved gates per Grover iteration depends on the physical
realization of the oracle function $f(x,y)$. 
Many template points (e.g., bold and large letters) 
lower the number of required Grover iterations.

Similarly, one has to estimate the failure probability, i.e., 
how often measurement of one of the ancilla bits yields zero.
The probability for the photons to be reflected (first ancilla) is 
given by the ratio of the image size (number of points) over the total
area of the array (number of pixels),  
i.e., many image points are favorable (similar to the above argument).
For the second ancilla the failure probability depends on the
amplitudes of the removed $k$-values and thus roughly on the amount of
noise and the number of fine details in the image. 

For specific templates we have made our algorithm explicit:
With a given template {\bf A} or {\bf B} 
(cf.~figure \ref{Fimage} top panels),  
we numerically calculated the recognition probability for all the
possible combinations 
(image {\bf A} and template {\bf B} versus template {\bf A} etc.) 
for various noise levels (cf.~figure \ref{Fimage} lower panels).

\begin{figure}[ht]
\includegraphics[width=8.5cm]{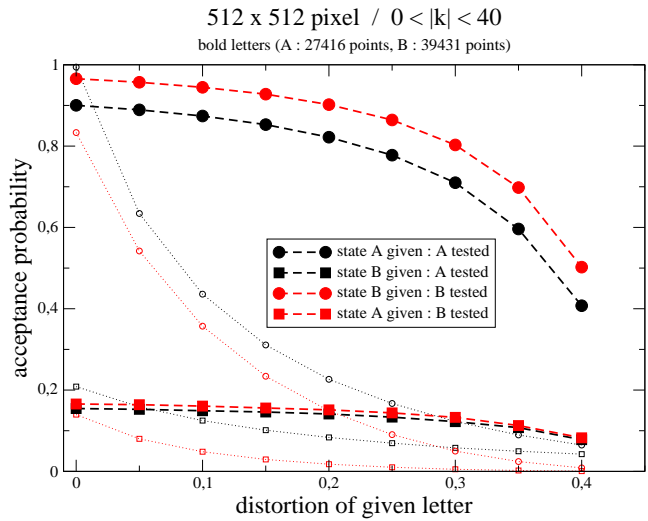}
\caption{\label{Fpatternmatching}
[Color Online]
Probability to recognize the initial state as the template versus
noise level.  
Small hollow symbols represent results without noise-filter applied. 
}
\end{figure}

The results of the numerical simulations are shown in figure
\ref{Fpatternmatching}.
The slight differences between the letters {\bf A} and {\bf B} result
from the different numbers of image points occupied by the two
templates (discretization error \cite{discretization} of the Grover
iterations etc.), as has been checked by applying the algorithm to a
different (more symmetric) font. 
An important quantity is the the discrimination ability of the
algorithm, i.e., the capability to distinguish between the
alternatives in the given alphabet {\bf A} and {\bf B}, see 
figure~\ref{Fpatternmatching}. 
A rough measure for the discrimination ability is given by the 
difference between the acceptance probability when the input state  
corresponds to the template (circle symbols) and the acceptance
probability when the input state does not match the template 
(square symbols).
Without noise, this difference is over 70\% and even with 40\% noise,
the discrimination ability is still significant (around 30\%),
provided that a noise filter is applied.
Without noise filter, the discrimination ability decreases much
faster.

\section{Failure probability}\label{Sdiscrimination} 

As already mentioned in sections \ref{Sproblem_definition} and 
\ref{Squantum_algorithm}, the task in section \ref{Squantum_algorithm}
is similar to the well-known discrimination problem:
Given a quantum state out of a set of perfectly known (non-orthogonal) 
quantum states, it is possible to construct a positive operator valued
measure (POVM) which decides which one of these states it is.
Since non-orthogonal states cannot be distinguished with certainty,
such a POVM measurement can never be performed without any errors or
ambiguities \cite{nielsen2000}. 
However, a POVM can be optimal in the sense that it minimizes the error
probability. 
Another optimal choice would be a POVM that never leads to an error
but sometimes to inconclusive results, but this choice is not suitable
for quantum states that are not known exactly. 

The explicit construction of the POVM operator has only
been possible for a few simple cases.
For example, if only the two states $\rho_A = \ket{A}\bra{A}$ and  
$\rho_B = \ket{B}\bra{B}$ can occur with probabilities $p_A+p_B=1$,
an optimal POVM can be constructed \cite{helstrom1976,bergou2004} via
the projectors on the eigenvectors of the operator
\bea
\label{Elambda}
\Lambda = p_A \rho_A - p_B \rho_B
\,.
\eea
Keeping in mind that the dimension of the Hilbert space grows
exponentially, finding the corresponding eigenvectors already 
becomes a computationally challenging task.
More important however is the difficulty that in an experiment, the
measurement of such an observable must be implemented. 

\begin{figure}[ht]
\includegraphics[width=8.5cm]{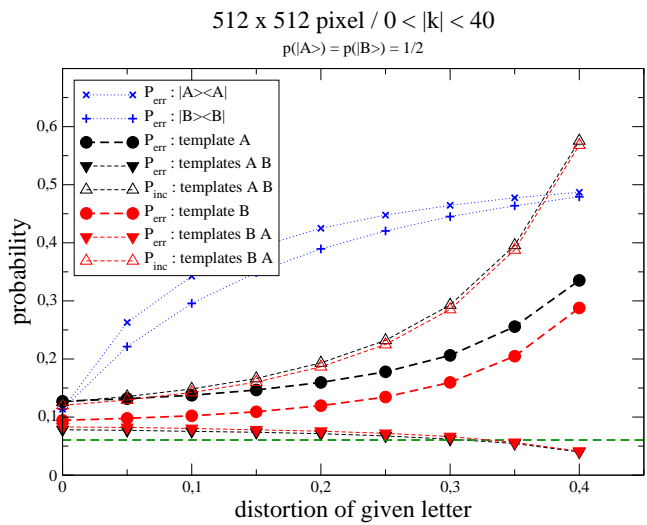}
\caption{\label{Ferror_prob_9}
[Color Online]
Error probability of the quantum algorithm when applied to state
discrimination for equally distributed states $p_A = p_B = 1/2$.
The dashed horizontal line corresponds to the minimum error
(\ref{Ehelstrom}) for unperturbed states $\ket{A}$, $\ket{B}$
(as in figure \ref{Fimage} top left panels).
The errors of the quantum algorithm (large circle symbols) correspond
to the averaged failure probabilities in figure \ref{Fpatternmatching}
with the respective template. 
For comparison, the error probabilities obtained from the expectation 
values of the projection operators $\ket{A}\bra{A}$ and
$\ket{B}\bra{B}$ have been added (small symbols). 
At the price of introducing an inconclusive result (hollow triangle
symbols), the probability of obtaining a conclusive but wrong result
can be lowered even further (filled triangle symbols), 
see section~\ref{Sae}.
Note that the POVM bound does not apply in this case and that the
decline of this error probability with increasing noise is due to the
fact that the probability of obtaining a conclusive result decreases
also.
The conditional probability of getting the wrong answer provided that
the algorithm arrived at a conclusive result of course increases with
noise -- but is still below the other curves (not shown).
}
\end{figure}

Associated with an optimal POVM one obtains the minimum error in 
distinguishing between the states $\ket{A}$ and $\ket{B}$
from the Helstrom formula \cite{helstrom1976}
\bea
\label{Ehelstrom}
P_{\rm err} = \frac{1}{2}\left(
1 - \sqrt{1 - 4 p_A p_B \abs{\bracket{A}{B}}^2}\right)
\,.
\eea
Of course, such a simple construction is only possible if
non-perturbed quantum states are used.
Nevertheless, one can compare the error probability of the quantum
algorithm in case of quantum state discrimination with the optimal
one, see figure \ref{Ferror_prob_9}. 
Without noise, the error probability of the quantum algorithm lies
only a few percent above the lower bound of about 6\%.
In the presence of noise, the quantum algorithm is of course more
likely to make mistakes, but then it should not be compared with the
POVM bound since this was derived assuming exactly known quantum
states. 
For comparison, we plotted the performance of the naive POVMs
containing the projector expectation values $\ket{A}\bra{A}$
and $\ket{B}\bra{B}$ (dotted lines), respectively, which are clearly
inferior to the presented quantum algorithm in the presence of noise.

As will be discussed in the following section, it is possible to
reduce the error probability even further at the expense of
introducing an inconclusive result. 

\section{Algorithmic Extensions}\label{Sae}

Note that the algorithm is vulnerable to translation, i.e., it does
not recognize an image that has been translated in position space
compared to the template.
To cope with translations, it is necessary to use more than one
reflected photon and to analyze the two-photon correlations.
After determining the center of mass of the image, for example, one
could perform the same algorithm with a template that has been shifted
accordingly. 

Since the objective is to shine as few photons as possible onto the
image, one should try to extract as much information as possible from
single measurements. 
Unfortunately, after a complete measurement of all the qubits
separately, the full quantum state has been projected onto 
the outcome and no information is left.
However, in order to accept or reject the hypothesis of a given
template, we do not need to measure all the qubits.
It is completely sufficient to ask the question ``Are {\em all} the
non-ancilla qubits in the state $\ket{0}$ or not?''
Alternatively, one could also construct a unitary operator 
(as with the noise-filter operator $\mathcal N$) that uses another
ancilla qubit to answer this question.
If the answer by measurement is ``yes'', we have accepted the
hypothesis (probabilistically) and the full quantum state has been
projected onto the state 
$\ket{0^{(n_x)}}\otimes\ket{0^{(n_y)}}\otimes\ket{1}\otimes\ket{1}$.  
However, if the answer is ``no'' and we have rejected the hypothesis 
(again probabilistically), the quantum state has only partially been
projected onto the high-dimensional sub-space orthogonal to 
$\ket{0^{(n_x)}}\otimes\ket{0^{(n_y)}}$ and still contains a
significant amount of information. 

For example, let us assume that we performed the algorithm with the 
template {\bf A} and that this hypothesis has been rejected, i.e., 
not all of the qubits were in the state $\ket{0}$.
One possibility for this rejection would be that the input
state was $\ket{B}$ or some other quantum state (rejection correct) or
that an input state $\ket{A}$ was falsely rejected (either due to
strong perturbations or  just due to the inherent probabilistic
nature, see figure \ref{Fpatternmatching}). 
In this case, we can partially undo the operations specific to
template {\bf A} 
by applying a Hadamard gate onto each qubit again and Grover-rotating
back with the template {\bf A}.
Now we may switch to the template {\bf B} and perform the inverse
Grover rotation with the new template {\bf B} plus the Hadamard gates 
and measure the outcome.
In case the overlap between the images {\bf A} and {\bf B} is small
enough, the result of this measurement indicates whether the image is
probably {\bf B} and the rejection of {\bf A} was correct or whether
something else happened (e.g., the rejection was erroneous or the
image is neither {\bf A} nor {\bf B}).

\begin{figure}[ht]
\includegraphics[width=2.05cm]{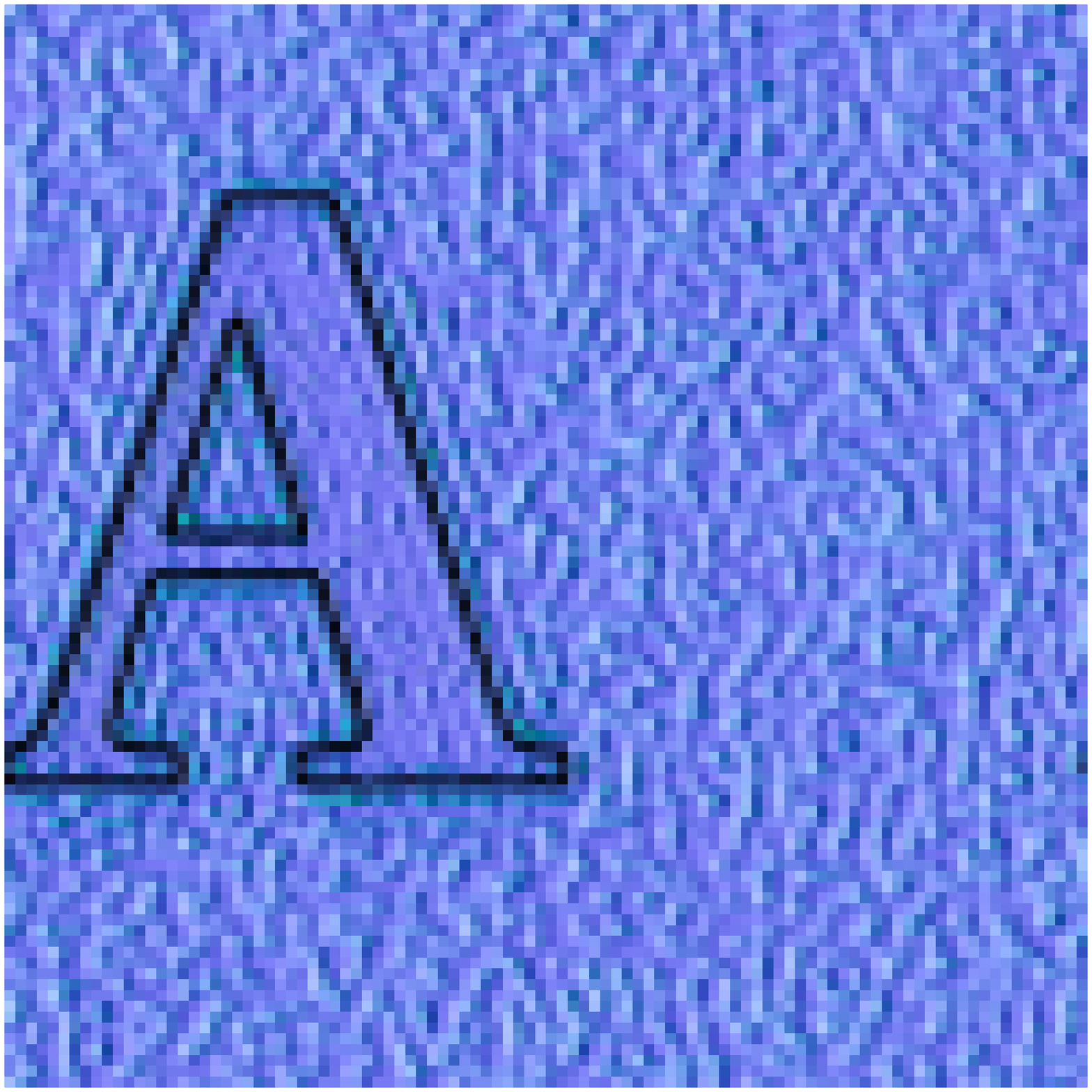}
\includegraphics[width=2.05cm]{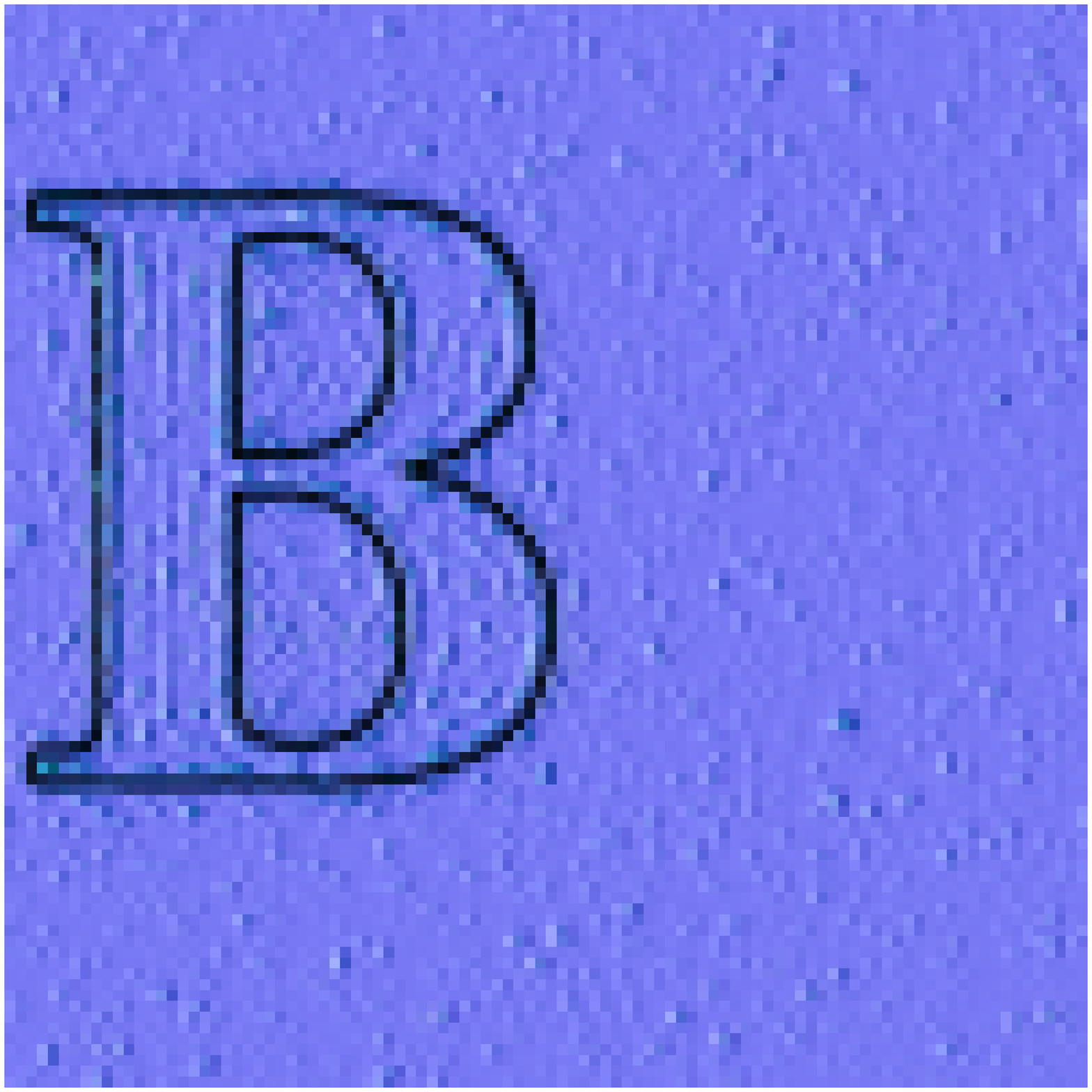}
\includegraphics[width=2.05cm]{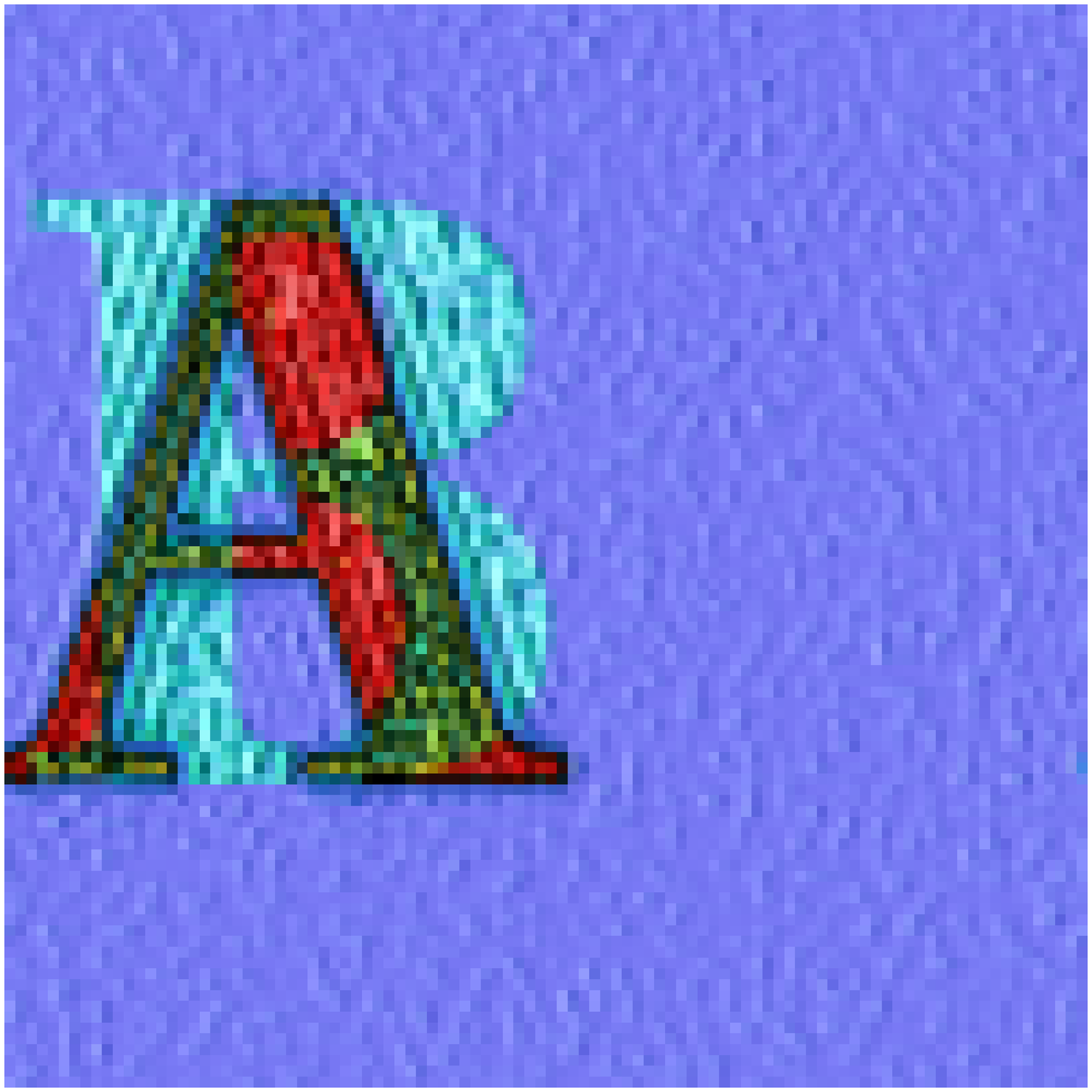}
\includegraphics[width=2.05cm]{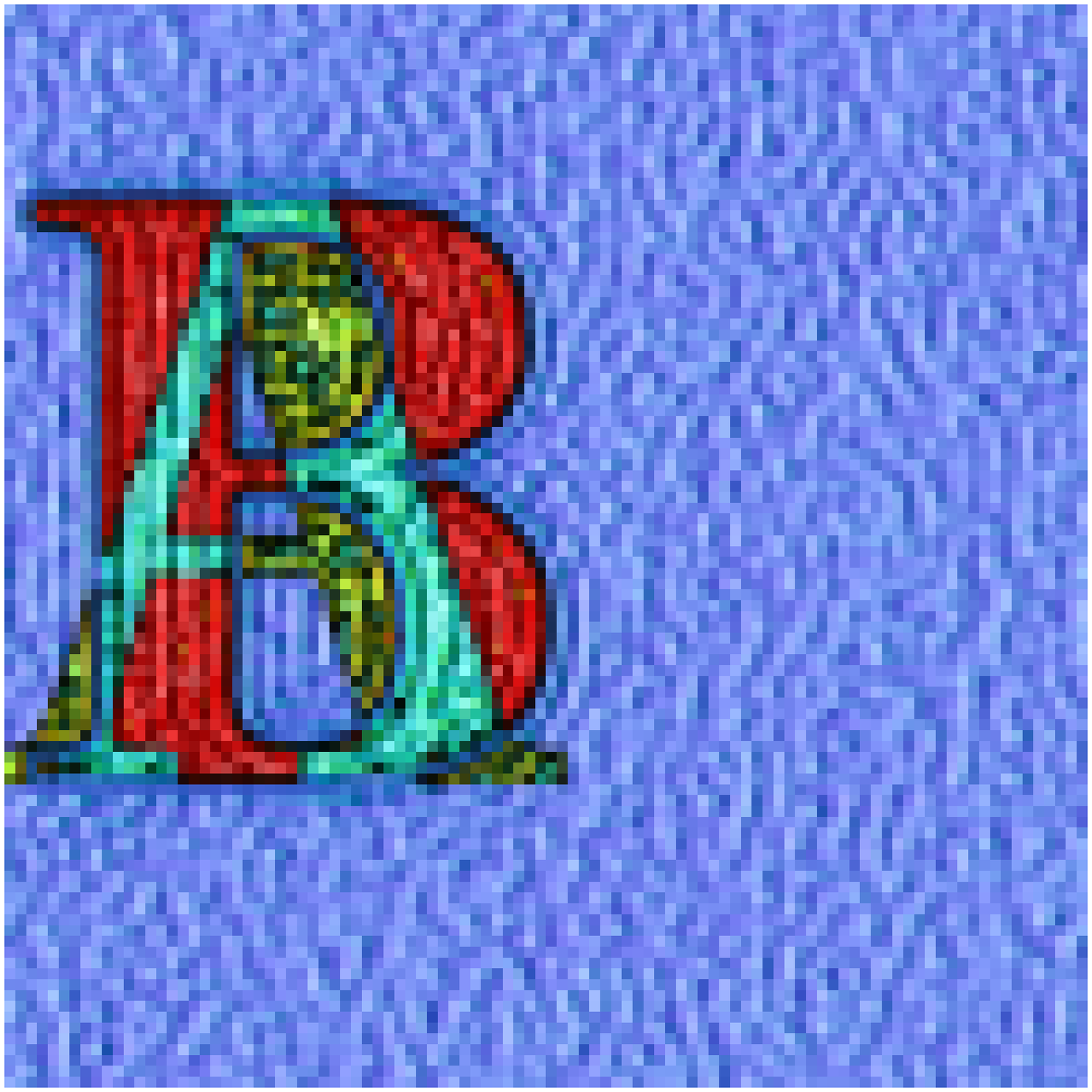}
\caption{\label{Fsecondtry}
[Color Online]
Color-coded distribution of the squared amplitudes of the state after
step 12 of the quantum algorithm (rejection of the first hypothesis). 
Color-coding has been chosen as in figure \ref{Fimage} and all images
have been perturbed by 5\% initially.
From left to right, the combinations have been chosen as follows: 
(image~{\bf A}~\&~template~{\bf A}), 
(image~{\bf B}~\&~template~{\bf B}), 
(image~{\bf A}~\&~template~{\bf B}), 
(image~{\bf B}~\&~template~{\bf A}).
}
\end{figure}

Thus, if the first template had $M_{\rm tp}^{(1)}$ points and the
second template had $M_{\rm tp}^{(2)}$ points, we could continue the
quantum algorithm as follows
\begin{itemize}
\item[11.] 
apply Hadamard gates on non-ancilla qubits\\
$\ket{\Psi_{11}}=
{\mathcal H}^{(n_x)}\otimes{\mathcal H}^{(n_y)}\ket{\Psi_{10}}$
\item[12.] 
perform 
$R=\left\lceil{\frac{\pi}{4}\sqrt{N_x N_y/M_{\rm tp}^{(1)}}}\right\rceil$ 
Grover iterations 
with respect to template {\bf A}\\
\\
$\ket{\Psi_{12}}={\mathcal G}^{R}_A\ket{\Psi_{11}}$
\item[13.] 
switch to the second template {\bf B} and perform \\
$R'=\left\lceil{\frac{\pi}{4}\sqrt{N_x N_y/M_{\rm tp}^{(2)}}}\right\rceil$ 
inverse Grover iterations with respect to template {\bf B}\\
\\
$\ket{\Psi_{13}}={\mathcal G}^{-R'}_B\ket{\Psi_{12}}$
\item[14.] 
apply Hadamard gates on non-ancilla qubits\\
$\ket{\Psi_{14}}=
{\mathcal H}^{(n_x)}\otimes{\mathcal H}^{(n_y)}\ket{\Psi_{13}}$
\item[15.] 
measure final state in computational basis\\
$\ket{\Psi_{15}}\stackrel{?}{=}
\ket{0^{(n_x)}}\otimes\ket{0^{(n_y)}}\otimes\ket{1}\otimes\ket{1}$
\end{itemize}
The measurement at step 10 of the quantum algorithm essentially
subtracts the template {\bf A} from the state. 
Consequently, in case of a false rejection, the state amplitudes will
concentrate on the perturbations of the image and the interference
patterns resulting from the Fourier noise-filtering, see figure
\ref{Fsecondtry} left panels. 
The resulting state is then nearly orthogonal to the second hypothesis
in general.  
In case of a correct rejection of the first hypothesis, the state's
squared amplitudes will decrease where the state overlaps with the
first template, see figure \ref{Fsecondtry} right panels. 
Note that here also the different magnitudes of the amplitudes due to
the different number of points will in  most cases prohibit a complete 
removal of these amplitudes.
In this case, the overlap between the state and the second template
will be substantial.

\begin{figure}[ht]
\includegraphics[width=8.5cm]{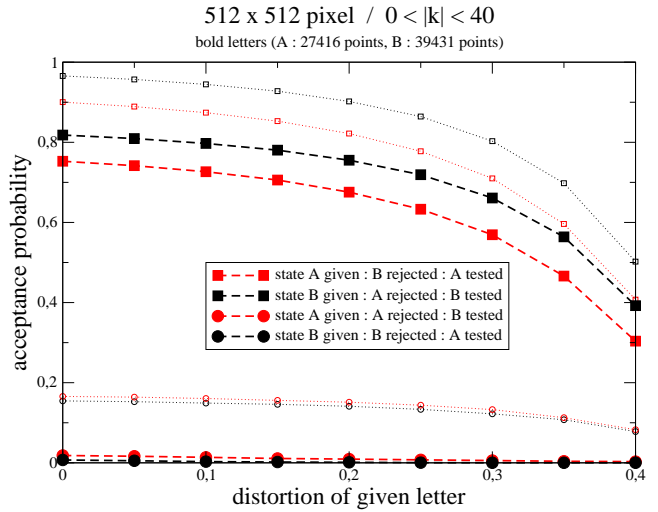}
\caption{\label{Fsecondtryprob}
[Color Online]
Probability for acceptance of the second hypothesis after rejection 
of the first one.
For comparison, the small hollow symbols (same data as large symbols
in figure \ref{Fpatternmatching}) -- corresponding to the acceptance
probability if the second hypothesis had been tested as the first one
-- have been included.}
\end{figure}

In this second attempt, the recognition probability for a correct
hypothesis is naturally a bit smaller as if the second hypothesis had
been tested directly, see large versus small symbols in figure 
\ref{Fsecondtryprob}. 
However, this procedure has the advantage, that no further interaction
with the image is necessary. 
A further important advantage is that the probability of falsely
recognizing a template is extremely low 
(large circle symbols in figure \ref{Fsecondtryprob}). 
This is due to the small overlap of the second hypothesis template and 
the falsely rejected state vector 
(compare again figure \ref{Fsecondtry} left panels).

With two templates given, one has only three different outcomes:
The first hypothesis can be accepted (1), the first hypothesis can be
rejected and the second one is then accepted (2), or none of the
hypotheses is accepted (3). 
In the last case, given that only $\ket{A}$ and $\ket{B}$
can occur, we know that an error must have happened.
Consequently, the last outcome must be regarded as inconclusive.
With the triangle symbols in figure \ref{Ferror_prob_9} it becomes
visible that with the extended algorithm, the error probability is
significantly lowered, whereas the probability for obtaining the
inconclusive result rises for strongly perturbed images 
as one would expect. 
Consequently, in alphabets containing only few letters, one can
identify the letter with a reasonable probability and detect false
rejections with high probability -- with just a single reflected
photon.  
Thus, the algorithmic extension is similar to a mixture of minimum
error and inconclusive POVMs -- with the advantage of being applicable
to perturbed quantum states.

\section{Discussion and Summary}

We have demonstrated that a quantum algorithm would be far more 
effective than a classical computer in template recognition -- firstly
regarding the number of incident photons and, secondly, in view of the
required number of gates. 

In case of a successful run, a single photon will suffice to identify 
a pattern (probabilistically) without disturbing the sensitive image. 
In this sense, the quantum algorithm realizes a destruction-free 
measurement (though probabilistically), which is similar to the
Elitzur-Vaidman problem \cite{elitzur1993}. 
Even when one takes the finite success probability of the quantum
algorithm into account, it still constitutes a major advantage 
in contrast to classical pattern recognition since a few incident
photons suffice.
In case of hypothesis rejection, the algorithm can be used to test a 
different hypothesis without necessitating a further photon
interacting with the sensitive array.
(Interestingly, the probability of falsely recognizing an input state 
as the template is much lower in these secondary trials than in the
first ones.)

The number of qubits required to run the quantum algorithm on a
$2^{n_x}\times 2^{n_y}$ image is $n_x+n_y+2$ and the number of
gates scales as $n_x^2+n_y^2$ for the Quantum Fourier transform, which
is also much faster than classical methods. 
The number of necessary Grover iterations will be rather
small \mbox{$\order{\sqrt{2^{n_x+n_y} M_{\rm tp}^{-1}}}=\order{1}$}, since for
reasonable images the number of points is comparable to the number of
pixels. 
Note that the measurements required during the quantum algorithm
can also be performed at the end. 

However, we have only discussed the algorithm from a theoretical point 
of view, neglecting many obstacles:
For example, we did not discuss the experimental difficulties that are 
to be expected during an experimental implementation of the required
quantum circuit. 
Apart from the quantum computer itself, the realization of the
quantum-controlled refractors poses serious problems.
Interestingly, for images of moderate size, the proposed quantum
algorithm can in principle be realized with present-day technology
using linear optics, see the Appendix.
However, the price one has to pay is an exponential scaling of the
number of gates etc.
Therefore, only the first advantage 
(i.e., only a few incident photons are required) of the presented
quantum algorithm survives in that case. 

Despite of these shortcomings, we hope that the presented theoretical
discussion of the amazing potential of quantum algorithms will further 
contribute to the theoretical as well as experimental developments in
this fascinating area. 
\section*{Note added in proof} 
If it is experimentally feasible to arrange to photon path is a way such that 
it interacts with the image several times, it is possible to increase the 
efficiency of the initial state preparation (i.e., suppress the probability 
for the photon being absorbed) by exploiting the quantum Zeno effect along 
the lines of P.~Kwiat {\em et al.}, Phys.\ Rev.\ Lett.\ {\bf 74}, 4763, (1995).

\section*{Acknowledgments}
%
This work was supported by the Emmy Noether Programme of the German
Research Foundation (DFG) under grant No.~SCHU~1557/1-1,2.

\section*{APPENDIX: LINEAR-OPTICS SETUP}

Instead of a binary representation, the quantum state can be encoded
directly by the path of a single photon in the laboratory.
This representation has the advantage that the proposed quantum
algorithm can be realized using linear optics elements which are
(in principle) already available with present-day technology, see also
\cite{londero2004}.  
Unfortunately, this advantage goes along with a serious drawback: 
The number of elements grows linearly (plus logarithmic corrections)
with the number of image pixels, i.e., the computational complexity is
exponential instead of polynomial.
Hence, this linear-optics scheme is only reasonable for images of
moderate size. 

\begin{figure}[ht]
\includegraphics[width=7cm]{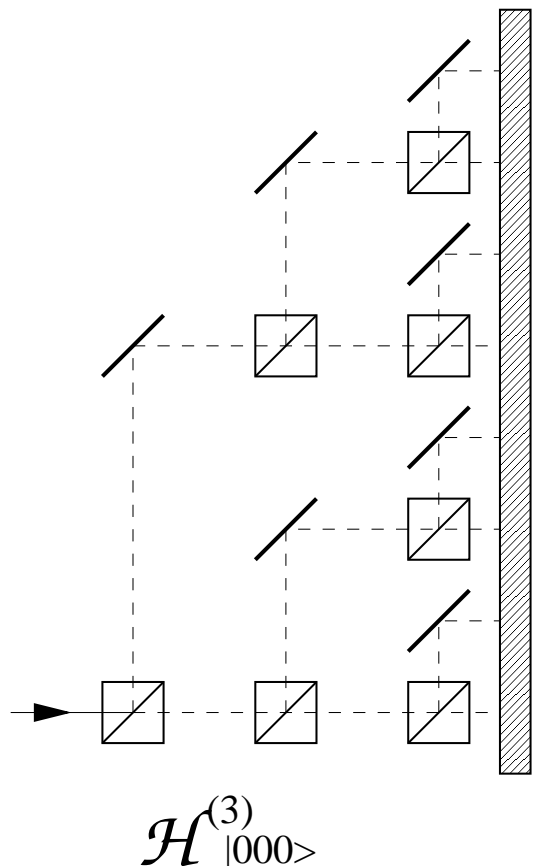}
\caption{\label{Fsplitter}
Preparation of the initial superposition state $\ket{s}$.\\
A Hadamard gate on the three-qubit-state $\ket{000}$ is emulated by
the combination of beamsplitters (crossed boxes) and mirrors (solid
lines), which distribute the amplitude of the incoming photon
uniformly on the one-dimensional array (large hatched box, partially
absorptive and transmittive).
}
\end{figure}

The initial state 
\mbox{$\ket{s}={\mathcal H}^{(n_x)}\ket{0^{(n_x)}}\otimes
{\mathcal H}^{(n_y)}\ket{0^{(n_y)}}$} 
can be generated by means of a series of beam splitters as depicted in 
figure \ref{Fsplitter} which distribute the photon amplitude uniformly
over the array and thus replace the quantum controlled refractors.
For a one-dimensional array of $N=2^n$ pixels, this scheme necessitates 
$N-1$ beam splitters.
%

For this implementation, it is convenient to regard the array
as being partially absorptive and transmittive.
If the photon passes the array, the quantum state is automatically 
prepared in a superposition of all points as in equation (\ref{points}).
On the other hand, if the photon is absorbed, no quantum state is
created at all, i.e., the existence of a photon corresponds to the
measurement of the first auxiliary qubit.

\begin{figure}[ht]
\includegraphics[width=8.5cm]{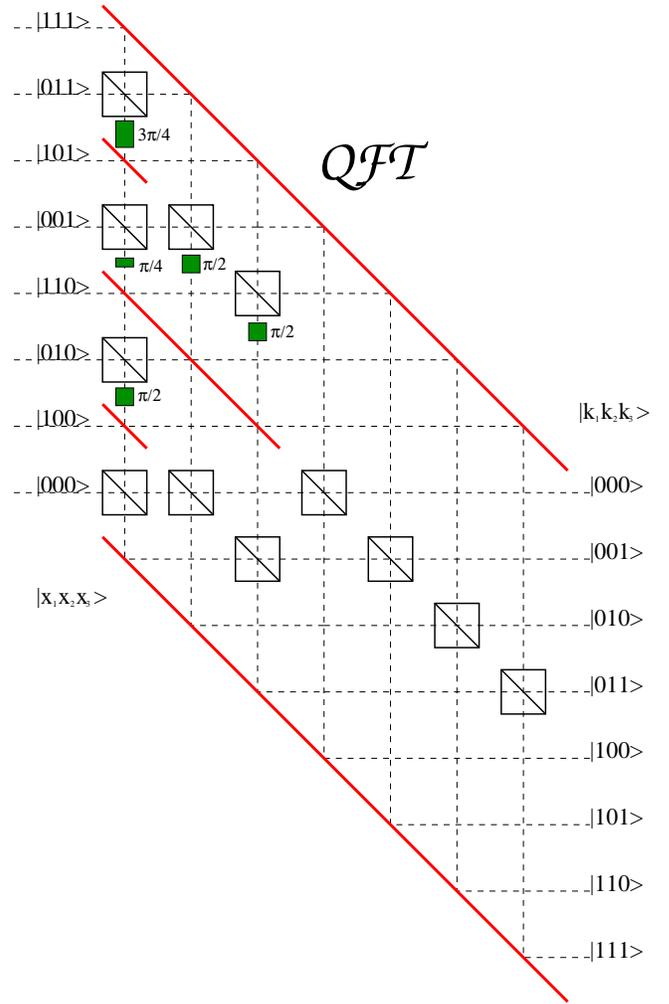}
\caption{\label{Fqft}
[Color Online]
Linear-optical setup for a quantum Fourier transform on a
8-dimensional Hilbert space corresponding to an array with 8 pixels. 
After interaction with the array, the quantum state (left-hand side) 
is described by the path of the photon (dashed lines). 
The quantum Fourier transform is performed using beamsplitters
(crossed boxes), mirrors (red solid lines) and phase shifters  
(green boxes) only. 
The distances between the beam paths are assumed to be a multiple of 
the photon wavelength, such that different distances in perpendicular 
direction do not contribute to the relative phase.
The beamsplitters are assumed to act with a phase factor of $-1$ if 
traversed vertically on a straight line 
(and without any phase otherwise).
The phase shifts (numbers beside boxes)
have been adjusted according to equation (\ref{Ephaseshift}).
This scheme only reaches the efficiency of the classical Fast Fourier
Transform (FFT).
Without the additional phase shifts, the scheme implements a Hadamard gate on all
qubits.
}
\end{figure}

Afterwards, a quantum Fourier transform can be performed with the
setup in figure \ref{Fqft}.
In the binary representation $k=\sum\limits_\ell k_\ell 2^{n-\ell}$
and $x=\sum\limits_j x_j 2^{n-j}$, the quantum Fourier transform,
\bea
{\cal QFT}\ket{x}
=
\frac{1}{\sqrt{N}} 
\sum_{k=0}^{N-1} 
\exp\left\{\frac{2 \pi i x k}{N}\right\} 
\ket{k}
\,,
\ea
possesses the well-known factorization 
\bea
{\cal QFT}\ket{x}
&=&
\sum_{k=0}^{N-1} 
\ket{k}
\prod_{\ell=1}^n \frac{(-1)^{k_\ell x_{n-\ell+1}}}{\sqrt{2}}
\times
\nn
&&
\times
\prod_{m=2}^\ell 
\exp\left\{2 \pi i k_\ell x_{n+m-\ell} 2^{-m}\right\}
\,.
\ea
For example, one observes for $n=3$ the following phase contributions
$\Delta \Phi$
\bea
\label{Ephaseshift}
\Delta\Phi_{\ell = 3} &=& (-1)^{k_3 x_1} 
\exp\left\{2 \pi i k_3 \left[x_2 2^{-2}
+ x_3 2^{-3}\right]\right\}\nn
\Delta\Phi_{\ell = 2} &=& (-1)^{k_2 x_2} 
\exp\left\{2 \pi i k_2 \left[x_3 2^{-2}\right]\right\}
\nn
\Delta\Phi_{\ell = 1} &=& (-1)^{k_1 x_3}\,.
\eea
By adjusting the beamsplitters such that they automatically induce
phase factors of $-1$ if (and only if) traversed vertically on a
straight line, we can generate the above phase contributions of $\pi$,
i.e., the factors of $(-1)$.
The remaining phases 
(in the above example $3\pi/4$, $\pi/2$, and $\pi/4$) can be
implemented by ordinary phase shifters placed in the corresponding
paths, cf.~figure~\ref{Fqft}.
Note that further mirrors could be used to bring the input states 
$\ket{x_1 x_2 x_3}$ in the order of the computational basis.
Without these additional phase shifters, the arrangement in 
figure~\ref{Fqft} would correspond to a Hadamard gate on every qubit
${\cal H}^{(n)}$.  
The setup requires $n 2^{n-1}=\ord(N\ln N)$ beamsplitters,
i.e., it does not admit the implementation of a quantum Fourier
transform in a complexity that is only polynomial in the number of
qubits but only reaches the efficiency of the classical Fast Fourier
Transform (FFT).

Having calculated the quantum Fourier transform, the undesired
$k$-values can be removed by simply placing photon absorbers into
photon paths that correspond to the respective $k$-values 
(compare steps 5 and 6 of the quantum algorithm). 
The inverse Fourier transformation can be performed in analogy to
figure \ref{Fqft}. 
The implementation of the inverse Grover rotations 
(step 8 of the quantum algorithm) is also rather simple by
means of linear optics:
The oracle, which inverts all states belonging to the template, can be
realized by placing a transparent glass template of appropriate
thickness ($\pi$-phase) into the optical path. 
The remaining inversion gate $\f{1}-\ket{s}\bra{s}$ with respect to
the equally weighted superposition state $\ket{s}$ can be implemented
by a similar phase gate sandwiched between two Hadamard gates 
${\cal H}^{(n)}$ \cite{nielsen2000}. 
The required phase gate has to supplement all states except $\ket{0}$
(which is left unaffected) with a phase factor of  $-1$, whereas the
Hadamard gates can be implemented as in figure \ref{Fqft} without the
additional phase shifters. 

By placing a single-photon detector in the optical path corresponding
to the state $\ket{0\ldots0}$ in the computational basis without
disturbing the other paths, it is 
even possible to perform the final measurement in such a way that the
subsequent operations in section \ref{Sae} are possible.
In summary, the presented quantum algorithm can be realized
experimentally with a few photons using linear optics elements -- 
but the effort (number of devices) scales exponentially instead of
polynomially. 

Note that, in case the wavelength of the used photon is appropriate,
an alternative scheme can be realized in analogy to optical
filtering:
A suitably designed lens translates transversal components of
wavenumbers (i.e., beam directions) into positions in the focal plane
and thus acts as a quantum Fourier transform, cf.~\cite{photonics}
In this case, the complexity is still exponential -- but somewhat 
hidden in the lens construction and the needed accuracy etc. 

\newpage

\end{document}